\documentclass[useAMS,usenatbib]{mn2e}

\usepackage{bm}
\usepackage{amsmath}

\usepackage{url}

\usepackage{ulem}
\usepackage[T1]{fontenc}
\usepackage{calligra}

\usepackage{graphicx}
\usepackage{dcolumn}
\usepackage{bm}

\usepackage{times}
\usepackage{verbatim}
\usepackage{graphicx}
\usepackage{graphics,epsfig}

\usepackage{theorem}
\usepackage{makeidx}
\usepackage{epic}
\usepackage{amssymb}
\usepackage{bbm}
\usepackage{xy}
\usepackage{cancel}

\usepackage{CJK}

\newbox\grsign \setbox\grsign=\hbox{$>$} \newdimen\grdimen
\grdimen=\ht\grsign
\newbox\simlessbox \newbox\simgreatbox
\setbox\simgreatbox=\hbox{\raise.5ex\hbox{$>$}\llap
     {\lower.5ex\hbox{$\sim$}}}\ht1=\grdimen\dp1=0pt
\setbox\simlessbox=\hbox{\raise.5ex\hbox{$<$}\llap
     {\lower.5ex\hbox{$\sim$}}}\ht2=\grdimen\dp2=0pt

\newbox\simppropto
\setbox\simppropto=\hbox{\raise.5ex\hbox{$\sim$}\llap
     {\lower.5ex\hbox{$\propto$}}}\ht2=\grdimen\dp2=0pt

\title[
Chirality, extended MHD statistics and solar wind turbulence]
  {
  Chirality, extended MHD statistics and solar wind turbulence}
\author[J.-Z. Zhu
]
  {Jian-Zhou Zhu
  \thanks{jz@SCCFIS.org}
  \\
  Su-Cheng Centre for Fundamental and Interdisciplinary Sciences, Gaochun, Nanjing 211316 China\\
}

\begin{document}


\pagerange{\pageref{firstpage}--\pageref{lastpage}} \pubyear{2017}

\maketitle

\label{firstpage}

\begin{abstract}
We unite the one-flow-dominated-state (OFDS) argument of \citet{MeyrandGaltierPRL12} with the one-chiral-sector-dominated-state (OCSCS) one \citep[][]{hydrochirality} to form a nonlinear extended-magnetohydrodynamics (XMHD) theory for the solar wind turbulence (SWT), \textbf{ranging from the MHD- to subproton-, and even to subelectron-scale regimes} \citep[modifying the theory of][]{AbdelhamidLingamMahajanAPJ16}. Degenerate chiral states in \citet{MiloshevichLingamMorrisonNJP17}'s XMHD absolute equilibria are exposed with helical representation, to offer the basis of replacing the linear wave (of infinitesimal or arbitrarily finite amplitudes) arguments of previous theories with OCSDS. Possible connection of the OFDS-plus-OCSDS theory with the local minimal-energy/stability principle is also discussed.
\end{abstract}


\section{Introduction}\label{sec:intr}
Solar wind turbulence (SWT) data present multiple scaling regimes, starting from a power law with the scaling exponent approximately $-1$ at the `outer scales' \citep[see, e.g.,][and references therein]{WicksHorburyChenSchekochihinMNRASL2010} to that around $-4$ down at the sub-electron scales \citep[e.g.,][and references therein]{Alexandrova13,Sahraoui13}. At the deep sub-ion and even sub-electron scales, kinetic effects \citep[e.g.,][]{PassotSulem2015} enter and explanations resorting to gyrokinetic turbulence theory \citep[see, e.g.,][and references therein]{KrommesARF12} may be more complete, however, though Kolmogorov phenomenology and tentative statistical absolute equilibrium analyses have been performed for electrostatic cases \citep{PlunkJFM10,gkaeOLD}, general solid calculations would be formidable for clear illumination. Fluid models, especially extended magnetohydrodynamics \citep[XMHD, see, e.g.,][and references therein]{MiloshevichLingamMorrisonNJP17}, with the ion and electron skin depths, $d_i$ and $d_e$, included to reflect the relevant physical ingredients, may present the major physics concerning the multiple scaling regimes, among others. 

Barotropic XMHD equations in Alfv\'enic units, for momentum and induction, besides the continuity (mass conservation) equation $\partial_t \rho = -\nabla \cdot (\rho \bm{u})$, read
\begin{eqnarray}
\partial_t \bm{u}  = -\nabla \left[\Pi +  u^2/2+(d_e\nabla \times \bm{B})^2/(2\rho^2)\right] +\nonumber\\
+  \bm{u} \times\left(\nabla \times \bm{u}\right)+ \left(\nabla \times\bm{B}\right) \times \bm{\mathcal{B}}/\rho, \label{eq:X1}
\end{eqnarray}
\begin{eqnarray}
\partial_t \bm{\mathcal{B}} = \nabla \times \left(\bm{u} \times \bm{\mathcal{B}} \right) - d_i \nabla \times \left[\left(\nabla \times\bm{B}\right) \times \bm{\mathcal{B}}/\rho\right] +\nonumber\\
+ d_e^2 \nabla \times \left[\left(\nabla \times \bm{B}\right) \times \left(\nabla \times \bm{u}\right)/\rho\right], \label{eq:X2}
\end{eqnarray}
with
\begin{equation} \label{eq:X3}
\nabla \cdot \bm{B}=\nabla \cdot \bm{\mathcal{B}}=0, \ \bm{\mathcal{B}}=\bm{B}+d_e^2\nabla \times (\nabla \times \bm{B}),
\end{equation}
and $\Pi$ the enthalpy
\citep{AbdelhamidLingamMahajanAPJ16}. We also introduce potential vectors $\nabla \times\bm{\mathcal{A}}=\bm{\mathcal{B}}$ and $\nabla\times\bm{A}=\bm{B}$ for later usage. Taking the limit of $d_i \to 0$, one arrives at the \textit{inertial MHD} \citep{KimuraMorrisonPoP14}, believed to be relevant for \textit{sub-electron} scales of the Earth's magnetosphere and solar wind plasmas; with $d_e \to 0$, the familiar Hall MHD \citep[e.g.,][]{MeyrandGaltierPRL12}; and, the ideal single-fluid MHD is formally recovered with both $d_e$ and $d_i$ taken to be zero.

Like in the formulation of the full two-fluid model, two invariant generalized helicities result from two frozen-in (dually, Lie-carried 2-form) vorticities \citep[][]{LingamMiloshevichMorrisonPLA16}. The generalized helicities are not only \textit{Casimir}s, but also quadratic invariant rugged with respect to Galerkin truncations (see later discussions), thus must play important dynamical roles. For example, helicity leads to one-chiral-sector-dominated states \citep[OCSDS:][]{hydrochirality} in turbulence, whose exposition requires helical-mode representation.

\section{`Chiroids' structure of XMHD}\label{sec:HDXMHD}
Helical(-mode) representation is intrinsic, applicable for any flow domain $\mathcal{D}$ and which corresponds to the expansion of the variable $\bm{v}$ into eigen/chandrasekhar-Kendall modes of the nondimensionalized curl operator $\mathcal{C}=(-\nabla^2)^{-1/2}\nabla \times$ \citep[][see the latter for a constructive definition of this operator for numerical experiment]{Moses71,ChenShanMontgomeryPRA90,CCE03}. For example, if working with incompressible $\bm{u}=\bm{u}^{\pm}$ (the compressible flows may be similarly treated as shown by \citep[][]{Moses71}; see also \citet{ZhuJFM16}), we can write the generalized Fourier expansion 
    $\bm{u}^s=\sum_n \hat{u}^s_n \bm{\phi}_n^s,~ \nabla \times \bm{\phi}_n^s=s\lambda_n \bm{\phi}_n^s ~(\text{Beltramity}), ~\lambda_n>0$
with $s=\pm$ representing the \textit{chirality}: Of course, here `chirality' is simply used to emphasize the left or right handedness of the motions, not the chirality operator, distinguished from the helicity one, in relativistic quantum mechanics. For the cyclic box domain $\mathcal{D}=[0,2\pi)^3$, we have the helical decomposition in the standard Fourier expansion:
\begin{equation}\label{eq:FourierHelical}
\bm{v}
=\sum_s \bm{v}^s=\sum_{\bm{k},s} \hat{\bm{v}}^s(\bm{k}) e^{\hat{i}\bm{k}\cdot \bm{r}}=\sum_{\bm{k},s} \hat{v}^s(\bm{k})\hat{\bm{h}}_s(\bm{k})e^{\hat{i}\bm{k}\cdot \bm{r}},
\end{equation}
with $\hat{i}^2=-1$ 
and 
the following properties: $\hat{i}\bm{k}\times \hat{\bm{h}}_s(\bm{k})=sk\hat{\bm{h}}_s(\bm{k})$, $\hat{\bm{h}}_s(\bm{-k})=\hat{\bm{h}}_s^*(\bm{k})=\hat{\bm{h}}_{-s}(\bm{k})$ and $\hat{\bm{h}}_{s_1}(\bm{k})\cdot\hat{\bm{h}}_{s_2}^*(\bm{k})=\delta_{s_1,s_2}$. [Every complex variable, including the pure unit imaginary number, wears a hat and its complex conjugate (c.c.) is indexed by ``*''.] 
Kelvin's nomenclature ``chiroid'' for a chiral molecular will be used. [We denote the chirality coming with $\bm{k}$ with $s_{\bm{k}}$
, and similarly for those with $\bm{p}$ and $\bm{q}$; also, some simplifications of the notations for the self-evident $\hat{\bm{v}}_{\bm{k}}^{s_{\bm{k}}}$s are made without ambiguity.]
The bases can be \citep{W92}
$\hat{\bm{h}}_s(\bm{k})=(s\hat{i}\bm{l}+\bm{l}\times\bm{k}/k)/(\sqrt{2}l)$,
with $\bm{l}$ being an arbitrary vector, with module $l$, perpendicular to $\bm{k}$.

We focus on incompressible XMHD with $\nabla \cdot \bm{u}=0$ and $\rho=1$ \citep[][]{AbdelhamidLingamMahajanAPJ16,MiloshevichLingamMorrisonNJP17}, and derive the equations \citep[in the symmetric form as that for neutral fluids by][]{W92}
\begin{eqnarray}
    \partial_t \hat{u}^{s_{\bm{k}}}=&&
    \sum_{\bm{k}=\bm{p}+\bm{q}}^{s_{\bm{p}},s_{\bm{q}}} \hat{C}_{\bm{kpq}} [(s_{\bm{q}}q-s_{\bm{p}}p)\hat{u}^{s_{\bm{p}}}\hat{u}^{s_{\bm{q}}}+ \nonumber\\
    &&+(\frac{s_{\bm{p}}p}{1+d_e^2p^2}-\frac{s_{\bm{q}}q}{1+d_e^2q^2})\hat{\mathcal{B}}^{s_{\bm{p}}}\hat{\mathcal{B}}^{s_{\bm{q}}}], \label{eq:cxmhd1}\\
    \partial_t \hat{\mathcal{B}}^{s_{\bm{k}}}=&&s_{\bm{k}}k\sum_{\bm{k}=\bm{p}+\bm{q}}
    ^{s_{\bm{p}},s_{\bm{q}}} \hat{C}_{\bm{kpq}} \{[(1-\frac{d_e^2s_{\bm{p}}ps_{\bm{q}}q}{1+d_e^2q^2})\hat{u}^{s_{\bm{p}}}\hat{\mathcal{B}}^{s_{\bm{q}}}- \nonumber\\
    &&-(1-\frac{d_e^2s_{\bm{q}}qs_{\bm{p}}p}{1+d_e^2p^2})\hat{\mathcal{B}}^{s_{\bm{p}}}\hat{u}^{s_{\bm{q}}}]+\nonumber\\
    &&+d_i[(\frac{s_{\bm{p}}p}{1+d_e^2p^2}-\frac{s_{\bm{q}}q}{1+d_e^2q^2})\hat{\mathcal{B}}^{s_{\bm{p}}}\hat{\mathcal{B}}^{s_{\bm{q}}}]\},\label{eq:cxmhd2} 
\end{eqnarray}
with $\hat{C}_{\bm{kpq}}=\hat{\bm{h}}_{s_{\bm{p}}} \times \hat{\bm{h}}_{s_{\bm{q}}} \cdot \hat{\bm{h}}^*_{s_{\bm{k}}}/2$. The above system can be written in a somewhat abstract form for the `chiroids', labeled by $c$s, denoting the real and imaginary parts of $\hat{v}^s(\bm{k})$s:
    $\dot{c}_n=\sum_{lm} C_{lmn}c_l c_m$,
where the couplers $C$s are of such symmetrical properties that lead to the \textit{Liouville theorem} $\sum_n \partial \dot{c}_n/\partial c_n=0$ and the following \textit{global conservation laws}
\begin{eqnarray}
\frac{d\mathcal{H}}{dt} =&&  \frac{d}{2dt}\int_{\mathcal{D}} [u^2+B^2+d_e^2(\nabla \times \bm{B})^2] d^3\bm{r} \nonumber\\
 =&& \frac{d}{2dt} \sum_{\bm{k},s} [|\hat{u}^s|^2+\frac{|\hat{\mathcal{B}}^s|^2}{(1+d_e^2k^2)}]=0,\\\label{eq:XH}
\frac{dH_M}{dt} =&&  \frac{d}{2dt}\int_{\mathcal{D}} [\bm{\mathcal{A}}\cdot\bm{\mathcal{B}}+d_e^2\nabla \times \bm{u}\cdot \bm{u}] d^3\bm{r} \nonumber\\
 =&& \frac{d}{2dt} \sum_{\bm{k},s} [skd_e^2|\hat{u}^s|^2+\frac{s|\hat{\mathcal{B}}^s|^2}{k}]=0,\\\label{eq:XHm}
\frac{dH_C}{dt} =&&  \frac{d}{2dt}\int_{\mathcal{D}} [2\bm{u}\cdot\bm{\mathcal{B}}+d_i\nabla \times \bm{u}\cdot \bm{u}] d^3\bm{r} \nonumber\\
 =&& \frac{d}{2dt} \sum_{\bm{k},s} [\hat{u}^{s*}\hat{\mathcal{B}}^s+c.c.+skd_i|\hat{u}^s|^2] =0,\label{eq:XHc}
\end{eqnarray}
where $\mathcal{H}$ is the Hamiltonian (energy), $H_M$ the generalized magnetic helicity and $H_C$ the generalized cross helicity.

\section{Chirality of XMHD and SWT}\label{sec:AEandT}

\subsection{Nonlinear chirality beyond linear cyclotron waves}\label{sec:ChiT}
Chirality in Hall MHD has been discovered and argued \citep{MeyrandGaltierPRL12} with the following alignments in the linearized wave dispersion relation \citep{SahraouiGaltierBelmontJPP07}:
\begin{eqnarray}
\hat{\bm u}(\bm k)=-\frac{B_0 k_\parallel}{\omega}\hat{\bm b}(\bm k),  ~~~~~
\hat{\bm a}(\bm k)
=\frac{s}{k}\hat{\bm b}(\bm k)=\frac{s\hat{i}\bm{k}\times\hat{\bm{a}}(\bm{k})}{k},\label{eq:Alignment}
\end{eqnarray}
where $\bm{B}=\bm{B}_0+\bm{b}$, $\bm{b}=\nabla\times\bm{a}$ and $B_0 k_\parallel=\bm{B}_0\cdot \bm{k}$.
With
\begin{eqnarray}
    \sigma_m =\frac{\hat{\bm a}(\bm k)\cdot \hat{\bm b}^\ast(\bm k)+c.c.
    }{2|\hat{\bm a}(\bm k)||\hat{\bm b}(\bm k)|},~~~~~
    \sigma_c =\frac{\hat{\bm u}(\bm k)\cdot\hat{\bm b}^\ast(\bm k)+c.c.
    }{2|\hat{\bm u}(\bm k)||\hat{\bm b}(\bm k)|}.\label{eq:sigma}
\end{eqnarray}
\citet{MeyrandGaltierPRL12} found with Eq. (\ref{eq:Alignment}) the \textit{magnetic polarization} $P_m(\bm{k})=\sigma_m\sigma_c=\pm 1$ (the Alfv\'{e}n ion cyclotron wave dispersion relation corresponds to $+1$, while the whistler wave mode $-1$.)
They studied Hall MHD in the vortex frozen-in form
\begin{eqnarray}
    \partial_t\bm\Omega_h=\nabla\times(\bm u_h\times\bm\Omega_h),~~~~~~~~~~~~(h=R,L)\label{eq:HallVF}
\end{eqnarray}
where, $\bm\Omega_R=\bm{B}, \bm u_R=\bm u-d_i\nabla\times \bm{B}$ and $\bm\Omega_L=\bm{B}+d_i\nabla\times \bm u, \bm u_L=\bm u$,
and let in Eq. (\ref{eq:HallVF}) the ion fluid speed $\bm{u}_L=\bm{u}=0$, making the Hall MHD equation degenerate to the electron magnetohydrodynamic (eMHD) equation
\begin{eqnarray}\label{eq:emhd}
    \partial_t\bm{B}=-d_i \nabla\times[(\nabla\times\bm{B})\times \bm{B}],
\end{eqnarray}
whose linear wave corresponds to $P_m=-1$. They further interpreted the $k^{-7/3}$-sector spectrum as the result of eMHD.
Similarly, the electron fluid speed was then set to zero, leading to the ion magnetohydrodynamics (iMHD)
\begin{eqnarray}\label{eq:imhd}
    \partial_t(1-d_i^2\Delta)\bm{B}=d_i \nabla\times[(\nabla\times\bm{B})\times(1-d_i^2\Delta) \bm{B}],
\end{eqnarray}
where ion speed $\bm u_i=d_i\nabla\times \bm{B}$. Now the linear wave has $P_m=1$, which was also used to identify the chirality of the simulated turbulence. Coupled iMHD and eMHD flows constitute Hall MHD.

Our main point is that the two extreme values $\pm 1$ for $P_m$ found in the above are for linear waves, which themselves may not be appropriate in the case of their simulation \textit{without} a guide field $\bm{B}_0$ (though large-scale magnetic fluctuations might be coarse-grained to be a local guide field, the situation would be much more complicated), and may not be appropriate for characterizing eMHD and iMHD turbulence.
But, the insight that \textit{the turbulence is dominated by the iMHD and eMHD flow respectively at subsequent scale regimes}, which may be termed \textit{one-flow-dominated state} (OFDS), is valuable and the proper nonlinear theory may be a combination of the \textit{one-chiral-sector-dominated state} \citep[OCSDS --- a state dominated by the `+' or the `-' chiral sector:][]{hydrochirality} with OFDS:
Having neither introduced the mean magnetic field (i.e, $\bm{B}$ may be just $\bm{b}$), nor linearized the models, we have for uni-chiral case, $\sigma_m=s=-\sigma_{c}$, i.e.,
\begin{eqnarray}
    P_m=\sigma_c\sigma_m=\frac{\sum\limits_s -skd_i|\hat{u}^s(\bm{k})|^2}{\sum\limits_s k d_i|\hat{u}^s(\bm{k})|^2} \cdot \frac{\sum\limits_s s k|\hat{a}^s(\bm k)|^2}{\sum\limits_s k|\hat{a}^s(\bm k)|^2}=-1,
\end{eqnarray}
for $\hat{u}^s_R=-skd_i\hat{b}^s$, with the L dynamics removed, as given in the last paragraph.
Similarly, with R dynamics removed, if we have only one chiral sector, $\sigma_m=s=\sigma_c$ and $P_m=1$. So, we see that a nonlinear theory with uni-chirality in the helical representation offers $P_m=\pm 1$.
Since the spectra presented in Fig. 3 of \citet{MeyrandGaltierPRL12} are from those fluctuations of $P_m>0.3$ and $P_m<-0.3$, the interpretation should be combined with the argument of OCSDS. [The Kolmogorovian local interaction argument and dimensional analysis, as applied by \citet{MeyrandGaltierPRL12} to get $k^{-7/3}$ and $k^{-11/3}$ spectra, may be carried over to the (sub)system limited to only one chiral sector.]

OCSDS is nearly Beltramian only in the mono-wavenumber situation, which leads to the (anti)alignment between $\bm{u}$ (of the dominate flow) and $\bm{B}$ in OFDS and to weak or vanishing quadratic terms. The usage of linear-wave argument is `popular' in plasma literatures, which, in many cases are not necessary; for instance, relevant to our notion, it appears to us that the interesting hodographs of \citet{HeTuMarschYaoApJL12} for chiral magnetic fluctuations \citep[see also][]{TelloniBrunoMNRAS16} may not necessarily be related to just linear waves. Note also that OCSDS requires non-vanishing (generalized) helicity \citep{hydrochirality}, whose origin/emergence should also be addressed.


\subsection{XMHD chiroids absolute equilibria}\label{sec:GTandAE}
In many realistic situations and especially in numerical experiments, only finite modes are kept. And, it is expected that those properties `rugged' after truncation are of special dynamical importance. For instance, with the (generalized) Fourier expansion for the XMHD dynamics of the frozen-in (generalized) vorticity $\bm{\Omega}_{\chi}=\nabla \times \bm{P}_\chi$, we have by the \textit{Plancherel theorem}
\begin{eqnarray}
  H_{\chi}^{\pm} = \int_{\mathcal{D}} \bm{\Omega}_{\chi}^{\pm}\cdot\bm{P}_{\chi}^{\pm}d^3\bm{r}=  \int_{\mathcal{D}} \pm (-\nabla^2)^{1/2}\bm{P}_{\chi}^{\pm}\cdot\bm{P}_{\chi}^{\pm}d^3\bm{r}\nonumber \\
  =\sum_n \pm \lambda_n |\hat{P}_{\chi n}^{\pm}|^2: ~ \chi=1,2; \label{eq:Plancherel}
\end{eqnarray}
$H_\chi=H^+_{\chi} + H^-_{\chi}$ being ideal invariants, it follows that their truncated (say, the Galerkin-truncation, with all modes with $n$ larger than some $N$ put to be $0$) versions are also conserved by the truncated dynamics \citep{k73}: There are different ways to see this fact as a result of the detailed conservation laws of the triadic interactions, and probably the most simple one is that when the $m$th mode of chirality $c$ is truncated ($\hat{P}_{\chi m}^c$ put to $0$), nothing adds to the existing $dH_\chi /dt=0$.
For such a Galerkin truncated system, the Liouville theorem ensures an ultimate invariant probability measures describing the phase flows at the absolute equilibrium (AE) state \citep{Lee52,k73}. The convenient and physical one is the canonical distribution $\sim \exp\{-\alpha \mathcal{H} -\beta H_M - \gamma H_C\}$ as used by \citet{MiloshevichLingamMorrisonNJP17}, closely following whose notations $a:= \alpha$, $b := \beta/k$, $c := \gamma$, $f:=k(\beta d_e^2 + \gamma d_i)$ and $d := \alpha/(1+k^2 d_e^2)$, we present the three-dimensional (3D) modal spectral densities \citep[following][and his notation]{k73} instead of those 1D ones collected over the $k$ shell:
\begin{eqnarray}
U_K^s(\textbf{k}) := \langle |\hat{u}^s|^2 \rangle/2 = (sd+b)/\Delta^s_X,\label{eq:UKs}\\
U_{\mathcal{B}}^s(\textbf{k}):= \langle |\hat{\mathcal{B}}^s|^2 \rangle/2 = (sf+a)/\Delta^s_X, \label{eq:UBs}\\
Q^s(\textbf{k}):= \langle \hat{u}^{s*}\hat{\mathcal{B}}^s+c.c. \rangle/2 = sc/\Delta^s_X, \label{eq:Qs}\\
U_{\mathcal{H}}^s(\textbf{k})
= U_K^s(\textbf{k})+U_{\mathcal{B}}^s(\textbf{k})/(1+k^2d_e^2), ~\label{eq:UHs}\\
Q_M^s(\textbf{k}) 
= sk d_e^2 U_K^s(\textbf{k})+s U_\mathcal{B}^s(\textbf{k})/k, \label{eq:QMs}\\
Q_C^s(\textbf{k}) = Q^s(\textbf{k})+s k d_i U_K^s(\textbf{k}), \label{eq:QCs}\\
U_{\bullet}(\textbf{k})=U_{\bullet}^+(\textbf{k})+U_{\bullet}^-(\textbf{k}), ~
Q_{\bullet}(\textbf{k})=Q_{\bullet}^+(\textbf{k})+Q_{\bullet}^-(\textbf{k}), \label{eq:UallQall}
\end{eqnarray}
with
$\Delta^s_X=fb+ad-c^2+s(ab+fd)$.
Thus we have exposed the `degenerate states', obtaining the finer structure of the AE spectra than \citet{MiloshevichLingamMorrisonNJP17} with whom  we can check the agreement by (\ref{eq:UallQall}). Note that the decomposition not only physically separates the two chiral sectors, but also exposes the `mirror symmetries' of the spectral, not to mention that now the denominator is of third order and that the problem can be analytically tracked down to each zeros (poles of the spectra): by ```mirror symmetries' of the spectral'' we mean the poles are of opposite signs. With the fact that the pole(s) should be positive ($k>0$), one can then figure out all possible shapes of the spectra for physical inference \citep{hydrochirality}.

An important remark is that we should not take our results only as the chiral decomposition, but also as the `purely helical' (i.e., uni-chiral at each $\bm{k}$) AE which may allow completely novel features, such as that with the new physically relevant temperature parameter(s), the `negative (energy) temperature' \citep{hydrochirality}, for homochiral Euler, among other possibilities \citep{ZhuPoF14}.
We may apply the chiroids' AE for finer analysis of plasmas dynamics and the SWT. For clarity, we will follow \citet{AbdelhamidLingamMahajanAPJ16} and \citet{MiloshevichLingamMorrisonNJP17} to discuss separate regimes respectively.

\subsubsection{$d_e \to 0$: Hall MHD OFDS+OCSDS}
Having already the basic AE analyses for two-fluid plasma model in \citet{hydrochirality}, it is not necessary for us to start over again and would be enough to just offer some pertinent remarks by referring to the relevant results of \citet{MiloshevichLingamMorrisonNJP17} and the SWT.
Actually, we had derived the AE spectra directly from the Hall MHD chiroids dynamics 
(2013, 
two-page abstract in the 14th European Turbulence Conference
) and they agree with the spectra reduced from the above XMHD ones:
	\begin{eqnarray}
		U_K^s = (\alpha k + s\beta)/\Delta_H,~
		U_B^s  = (\alpha k + s\gamma  d_i)k^2/\Delta_H,~\label{eq:UKinHUmagH}\\
		 Q_M^s =  \frac{\gamma d_ik+s \alpha}{\Delta_H},~
	  Q_C^s =  \frac{sd_i\alpha k^2+\beta d_i k-2\gamma k}{\Delta_H}.\label{eq:QMHQCH}
	\end{eqnarray}
with $\Delta_H=s\alpha\gamma d_i k^2+(\alpha^2+\beta\gamma d_i-\gamma^2)k+s\alpha \beta$.
The spectra expose the `degenerate states' in and sum up to those of \citet{ServidioETC08}. Examining Eq. (\ref{eq:QMHQCH}), we can see that the concentration of the spectra in Fig. 2 of \citet{MiloshevichLingamMorrisonNJP17} is due to a pole $k_p$ slightly larger than their $k_{max}$. For appropriate $\alpha$, $\beta$ and $\gamma$, there can be two positive poles making $\Delta_H=0$ in Eqs. (\ref{eq:UKinHUmagH}) and (\ref{eq:QMHQCH}). Note that the poles of the two chiral sectors are of opposite signs as can be seen from the symmetry of the formula for the spectra. So, whatever cascade is inferred from the concentration of the spectra while approaching the pole(s), accompanying OCSDS could also be argued, which was most clearly demonstrated in \citet{hydrochirality} more explicitly and directly for the classical single-fluid MHD \citep{FrischETCjfm75}, whose results are just the above ones with $d_i=0$ (classical single-fluid MHD has only one flow, thus OFDS \textit{per se}.)

\citet{MeyrandGaltierPRL12} conjectured that ``the total (L + R) magnetic fluctuations spectrum should scale in $k^{-11/3}$ at large dispersive scales and in $k^{-7/3}$ at small dispersive scales'', and \citet{Sahraoui10} indeed showed, for the Cluster data, perpendicular magnetic power spectra a first steeper and then shallower spectra at subproton scales roughly corresponding to the Hall MHD regime, somehow supporting the OFDS-plus-OCSDS theory. For OCSDS in the respective iMHD and eMHD dominated regimes, the (generalized) helicity should be nonvanishing; otherwise, if we take $\beta=\gamma=0$ in the above, we are left with no pole to support OCSDS. But, note that, with finite mass, iMHD and eMHD are formally the same except for the opposite signs of the velocities carrying the frozen-in generalized vorticities [$\bm{\Omega}_{\chi}=\nabla \times \bm{P}_{\chi}$ with $\bm{P}_{\chi}=m_{\chi}\bm{u}_{\chi}+q_{\chi}\bm{A}$, and $\bm{u}_{\chi}=q_{\chi}\nabla \times \bm{b}$ written explicitly in terms of the physical mass $m_{\chi}$ and electric charge $q_{\chi}$ as in \citet{hydrochirality} for the two-fluid model, when the flow of one species is removed], and that \citet{MeyrandGaltierPRL12} explicitly stated that their forcing terms ``are chosen such as injection rates of cross helicity and magnetic helicity are null'', which, though without mentioning the kinetic helicity as part of the generalized helicity, indeed indicates null injection of the generalized helicity; thus, probably `spontaneous chirality' does happen in their simulation with the exchange of opposite-sign generalized helicity between iMHD and eMHD, though this `spontaneous chirality' is slightly different to what they originally meant. For SWT, besides the exchange of (generalized) helicities between iMHD and eMHD, there are other `external' mechanisms \citep{M78} which may help.

\subsubsection{$d_i\to 0$: inertial MHD OFDS+OCSDS}\label{sec:inertialMHD}
The first of \citet{MiloshevichLingamMorrisonNJP17}'s Eqs. (69--71) reads in our notations for the 3D modal spectral density
\begin{equation}\label{eq:UIMHD}
    U^{\pm} = 2/(\alpha\pm \gamma d_e k),
\end{equation}
which is consistent with what \citet{hydrochirality} observed in their Eq. (2.7) for deep-sub-electron scales of eMHD: Such consistency, with quantitative difference though, is not surprising, because both are for scales much smaller than $d_e$, and indeed \citet{Keramidas-CharidacosLingamMorrisonWhiteWurmPoP14} have demonstrated how XMHD is reduced to eMHD with the immobile-ion assumption (which is \textit{ad hoc}: P. Morrison, Private communication). [As \citet{hydrochirality} also observed, the deep-sub-electron-limit spectrum is exactly in the same form of neutral fluid, which in general can only have the large-$k$ pole indicating forward cascade and secondary OCSDS, and \citet{MiloshevichLingamMorrisonNJP17} shows the qualitatively consistent spectra in their Fig. 3.] \citet{AbdelhamidLingamMahajanAPJ16} argued that XMHD may be better than eMHD at the sub-electron regime to account the SWT and the interplanetary magnetic field \citep[IMF,][presenting actually an exponent $-4.2228\pm 0.011$, somewhat closer to $-13/4$]{LeamonSmithNessMatthaeusWongJGR98} with roughly a $k^{-4}$ spectrum \citep{Alexandrova13,Sahraoui13}.

In \citet{AbdelhamidLingamMahajanAPJ16}'s phenomenology for the sub-electron solar wind or interplanetary magnetic field, the alignment between $\hat{\bm{u}}_{\bm{k}}$ and $\hat{\bm{b}}_{\bm{k}}$ was invoked [for their Eq. (39), say], which, like \citet{MeyrandGaltierPRL12} is a result of the linear wave dispersion relation; but the new argument was that the relation is also a result of vanishing quadratic interaction terms [such modes can be seen directly from Eqs. (\ref{eq:cxmhd1}, \ref{eq:cxmhd2}) as other plasma fluid models \citep{ZhuExact}], thus the wave can be nontrivially of arbitrary finite amplitude. One way to think about it is to take XMHD turbulence as a superposition of such exact modes, which, by vanishing interactions, however would lead to vanishing spectral transfer, not the case of turbulence. An alternative argument is actually again `OFDS+OCSDS': The two (Lie-)carrying velocities for the generalized vorticities are $\bm{u}_{\pm}=\bm{u}-\kappa_{\mp}\nabla \times \bm{B}$ with $\kappa_{\pm}$ solving $\kappa^2 - d_i \kappa - d_e^2=0$ \citep{LingamMiloshevichMorrisonPLA16}, which indicates, in the inertial MHD limit, with $d_i=0$, $\bm{u}_\pm=\bm{u} \pm d_e \nabla\times\bm{B}$. Thus, we have $\hat{\bm{u}}_{\bm{k}}=\mp d_e \hat{i}\bm{k}\times \hat{\bm{B}}_{\bm{k}}$ by removing one of the two flows, which more realistically should be the dominance by one flow; that is, $\bm{u}_\pm  \thickapprox \pm 2d_e\nabla\times\bm{B}$ dominates with $\bm{u}_\mp \thickapprox 0$ (plasma physics considerations may indicate that $\bm{u}_-$ should dominate in this regime, which does not matter for our derivation now). Again, uni-chirality, or more realistically, OCSDS means $\hat{i}\bm{k}\times\hat{\bm{B}}_{\bm{k}} \thickapprox sk \hat{\bm{B}}_{\bm{k}}$, with $s=\pm$ depending on the chirality, so that $\hat{\bm{u}}_{\bm{k}}=- sk d_e \hat{\bm{B}}_{\bm{k}}$. Such $\hat{\bm{u}}_{\bm{k}}\propto k\hat{\bm{B}}$ scaling is exactly what \citet[][]{AbdelhamidLingamMahajanAPJ16} used in their Kolmogorov phenomehology to obtain $k^{-13/3}$. In summary, replacement of linear, finitely arbitrary-amplitude, wave argument with nonlinear OCSDS, united with OFDS, constitutes the basis of the inertial MHD turbulence cascade.

\begin{figure}
          \begin{center}
          \includegraphics[width=0.5\textwidth]{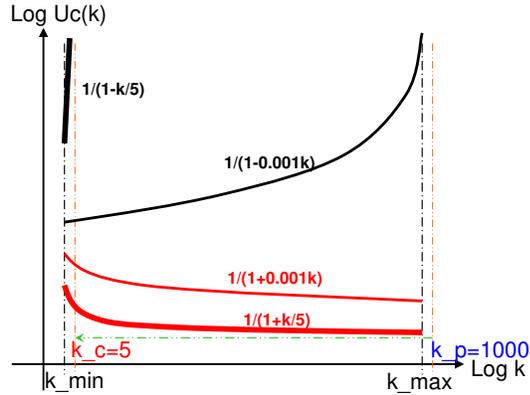}\\
          \caption{Schematic for $U^s(k)=1/(\alpha+s\beta k)$: $0<k_{min}< k_c=1$ and $k_{max}\gg k_c$. When $\alpha/\beta=1/0.001$ (thinner lines), the negative chiral sector pole $k_p=1000\gtrsim k_{max}$. This pole approaches $k_c$ when $\alpha/\beta \to 1/0.2$ (thicker lines) and most of the energy is concentrated at the negative chiral sector around the pole $k_p \to k_c=5$, the very small wavenumber compared to $k_{max}$: $k_c$ is the maximum wavenumber of the positive chiral sector, so the (dominant) positive helicity has opposite sign to the larger-wavenumber modes where injection is supposed be placed.}\label{fig:ocsds}
          \end{center}
\end{figure}

OFDS should be the natural result of `plasma physics' characterized by the scale, the mass (ratio, between ion and electron, say) etc., while OCSDS is a result of nonlinear helical thermalization/interaction dynamics.
However, at even smaller scales below the observed $k^{-4}$ regime, some dissipation mechanism may exist \citep[even for `collisionless' plasmas with `huge' mean free path, the `dissipation' might be caused, say, by the electron Landau damping: see, e.g.,][ for a recent discussion of SWT dissipation.]{SchreinerSaurAPJ17} If this indeed is the case, and, as pointed out in the above, for the neutral-fluid-like character of the spectra in the this regime, the dissipation mechanism may seriously destroy OCSDS, leaving the so-called second order OCSDS with only the more persisting cascade flux of one chiral sector \citep{hydrochirality}. Then we would need to find some other possibility to serve as the candidate of OCSDS mechanism, as is given below.
We use the idea and physical scenario explained by Eq. (2.12) of \citet{hydrochirality}, but now, for plasmas relevance, we further develop it explicitly with a diagram given in Fig. \ref{fig:ocsds}. Besides the caption, we still provide an alternative description: $U^c$ is the modal spectral density of the energy of neutral fluid, electron MHD or inertial MHD with $kd_e>>1$ (and $\beta=0$ or $\gamma=0$); and, originally both chiral sectors are left with $k_{min}<k<k_{max}$, while modes with $k\ge k_c$ of one of the chiral sectors are further truncated, allowing the pole to locate at some small $k_c$ around which the energy is concentrated. That is, the energy can be concentrated at some very-large but not the largest scales, which might persist in some turbulence state (since now the concentration scales are far from the turbulence dissipation scales: This is obvious for neutral fluids, and for plasma we now assume dissipation, if any, only takes place at scales smaller than the observed $k^{-4}$ sub-electron scaling range).
Of course, the above truncation scheme being so stringent, it may not appear very plausible to apply it to realistic flows
\citep[see more descriptions in][]{hydrochirality}.
Nevertheless, we have yet another ideal mechanism of OCSDS, and given the similarities between eMHD and inertial MHD at sub-electron scales with the neutral fluid, it may not be impossible that some mechanism or technique could somehow make it work (to `freeze' modes of one chiral sector); see more in the next section.

\section{Further discussions}
As other works \citep[e.g.,][]{AbdelhamidLingamMahajanAPJ16}, our theory has limitations in or has not yet touched those aspects discussed by \citet{PassotSulem2015} on Landau damping, \citet{ServidioETAL2015} on coherent structures and intermittency and \citet{BrunoCarbone2013} on parallel v.s. perpendicular magnetic fluctuations. But, since the OFDS-plus-OCSDS theory is fundamentally relevant in the scaling regimes from classical MHD to Hall MHD and to inertial MHD, we further suggest it make sense to set up different asymmetrical intermediate truncation wavenumbers (such as the $k_c$ in Fig. \ref{fig:ocsds}) at places \citep[say, those indicated by the vertical lines in the figures of][]{AbdelhamidLingamMahajanAPJ16} of the ion and electron skin depths or gyroradii, i.e.,
\begin{equation}
k_{ci}\simeq 1/d_i, ~k_{ce}\simeq 1/d_e.
\end{equation}
And, such asymmetrical truncations scheme, besides serving as the theoretical model, can also be applied to numerical simulations of XMHD for particular purposes of \textit{in cilico} experiments.

And, we remark that the connection between the AE calculation and minimum-energy state (MES) had been established for classical MHD \citep[][and references therein]{StriblingMatthaeusPoF91}, and the rapid and local relaxation leading to spontaneous emergence of chirality and Alfv\'enicity in spatial patches, has been argued and measured in numerical MHD turbulence and SWT \citep[][]{ServidioMattheausDmitrukPRL08,OsmanETCapj11}. Now, computing $\delta I/\delta \bm{u} = \textbf{0} = \delta I/\delta \bm{A}$ with $I = \mathcal{H} + \zeta \mathcal{H}_M + \xi \mathcal{H}_C$ to minimize $\mathcal{H}$ by introducing the Lagrangian multipliers $\zeta$ and $\xi$ \citep{Woltjer59}, we obtain
\begin{eqnarray}\label{eq:XMHDextreme}
        \zeta (\bm{\mathcal{B}} + d_e^2 \nabla \times \nabla \times \bm{\mathcal{B}}) + \xi (\nabla
        \times \bm{u} + d_e^2 \nabla \times \nabla \times \nabla \times \bm{u}) \nonumber \\
        + \nabla \times \bm{\mathcal{B}} = 0 = \bm{u} + (\zeta d_e^2 + \xi d_i)\nabla \times \bm{u} + \xi \bm{\mathcal{B}}.
\end{eqnarray}
It is seen that the OFDS-plus-OCSDS theory in Secs. \ref{sec:ChiT} and \ref{sec:inertialMHD} offer different possible ways to (approximately) realize (\ref{eq:XMHDextreme}), if locally, for different spatial patches characterized by correspondingly different $k$s (to form a spectrum): In each patch, OCSDS for a corresponding single characteristic $k_\star$ or Beltramity for the single dominant-excitation scale effectively make Eq. (\ref{eq:XMHDextreme}) agree with the relations of the OFDS in those regimes, including but not limited to the simple (anti)alignments between $\bm{u}$ and $\bm{B}$ in classical MHD [which, if indeed happens to SWT, then to some degree would reconcile locally with the linear-`wave' (due to, say, a local time-fluctuating guide field and/or rotation of time scale longer than that of the wave but shorter than the life of the patch) argument\textbf{, while} the nonlinear relaxation processes may somehow already be OFDS and OCSDS]. \textbf{And, it turns out possible to show that the} relaxed state admitted by (\ref{eq:XMHDextreme}) can lead to Beltramity and Alfv\'enicity, and, thus OFDS and OCSDS\textbf{: The details will be communicated elsewhere, with similar results holding also for the two-fluid model}. Actually, other variational principles may also produce similar chirality emergence features, even in different geometries \citep[][and references therein: here, say, minimal energy dissipation with given $\mathcal{H}$, $\mathcal{H}_M$ and $\mathcal{H}_c$]{Woltjer59}, calling for direct comparative observations and computations of the coherent patches and a unified treatment for SWT in particular, and plasma relaxation and dynamo in general.

\section*{Acknowledgments}
\begin{CJK*}{GB}{gbsn}
Supported by Ti\'an-Yu\'an-Xu\'e-P\`ai 
Foundation and NSFC \# 11672102. I thank Z. Guo and P.J. Morrison for discussions, and an anonymous referee for suggesting possible MES relevance.
\end{CJK*}


\begin{thebibliography}{99}

\bibitem[Abdelhamid, Lingam \& Mahajan (2016)]{AbdelhamidLingamMahajanAPJ16}
Abdelhamid H. M., Lingam M., Mahajan S. M., 2016, ApJ, 829, 87


\bibitem[\protect\citeauthoryear{Alexandrova et al.}{2013}]{Alexandrova13}
Alexandrova O. et al., 2013, Space Sci. Rev., 178, 101

\bibitem[\protect\citeauthoryear{Bruno \& Carbone}{2013}]{BrunoCarbone2013}
Bruno R., Carbone, V., 2013, Living Rev. Solar Phys., 10, 2

\bibitem[Chen, Shan \& Montgomery (1990)]{ChenShanMontgomeryPRA90}
Chen H., Shan X., Montgomery D., 1990, Phys. Rev. A, 42, 6158

\bibitem[Chen, Chen \& Eyink (2003)]{CCE03}
Chen Q. N., Chen S. Y., Eyink G. L., 2003, 
Phys. Fluids, 15, 361

\bibitem[\protect\citeauthoryear{Frisch et al.}{1975}]{FrischETCjfm75}
Frisch U. et al., 1975, J. Fluid Mech., 68, 769

\bibitem[He, Tu, Marsch \& Yao (2012)]{HeTuMarschYaoApJL12}
He J., Tu C-Y., Marsch E., Yao S., 2012, ApJ, 745, L8


\bibitem[Keramidas-Charidacos et al. (2014)]{Keramidas-CharidacosLingamMorrisonWhiteWurmPoP14}
Keramidas Charidakos I. et al., 2014, Phys. Plasmas, 21, 092118

\bibitem[Kim \& Cho (2015)]{KimChoAPJ15}
Kim H., Cho J., 2015, 
ApJ., 801, 75

\bibitem[\protect\citeauthoryear{Kimura \& Morrison}{2014}]{KimuraMorrisonPoP14}
Kimura K., Morrison P.J., 2014, Phys. Plasmas, 21, 082101

\bibitem[\protect\citeauthoryear{Kraichnan}{1973}]{k73}
Kraichnan R.H., 1973, J. Fluid Mech., 59, 745

\bibitem[Krommes (2012)]{KrommesARF12}
Krommes J. A., 2012, Annu. Rev. Fluid Mech., 44, 175

\bibitem[Leamon et al. (1998)]{LeamonSmithNessMatthaeusWongJGR98}
Leamon R. J. et al., 1998, J. Geophys. Res., 103, 4775

\bibitem[Lee(1952)]{Lee52}
Lee T.-D., 1952, 
Q. Appl. Math., 10, 69

\bibitem[Lingam, Miloshevich \& Morrison (2016)]{LingamMiloshevichMorrisonPLA16}
Lingam M., Miloshevich G., Morrison P.J., 2016, Phys. Lett. A, 380, 2400

\bibitem[\protect\citeauthoryear{Meyrand \& Galtier}{2012}]{MeyrandGaltierPRL12}
Meyrand M., Galtier S., 2012, 
Phys. Rev. Lett., 109, 194501


\bibitem[Miloshevich, Lingam \& Morrison (2017)]{MiloshevichLingamMorrisonNJP17}
Miloshevich G., Lingam M., Morrison P. J., 2017, New J. Phys., 19, 015007


\bibitem[\protect\citeauthoryear{Moffatt}{1978}]{M78}
Moffatt H.K., 1978, Magnetic Field Generation in Electrically Conducting Fluids, Cambridge University Press


\bibitem[\protect\citeauthoryear{Moses}{1971}]{Moses71}
Moses H.E., 1971, 
SIAM J. Appl. Maths, 21, 114


\bibitem[Osman et al. (2011)]{OsmanETCapj11}
Osman K. T. et al., 2011, ApJ, 741, 75

\bibitem[\protect\citeauthoryear{Passot \& Sulem}{2015}]{PassotSulem2015}
Passot T., Sulem P.L., 2015, ApJ, 812, L37

\bibitem[Plunk et al. (2010)]{PlunkJFM10}
Plunk G. G. et al., 2010, J. Fluid Mech., 664, 407

\bibitem[\protect\citeauthoryear{Sahraoui, Galtier \& Belmont}{2007}]{SahraouiGaltierBelmontJPP07}
Sahraoui F., Galtier S., Belmont G., 2007, J. Plasma Phys., 73, 723


\bibitem[\protect\citeauthoryear{Sahraoui et al.}{2013}]{Sahraoui13}
Sahraoui F. et al., 2013, ApJ, 777, 15

\bibitem[Sahraoui et al. (2010)]{Sahraoui10}
Sahraoui F. et al., 2010, Phys. Rev. Lett., 105, 131101


\bibitem[Schreiner \& Saur (2017)]{SchreinerSaurAPJ17}
Schreiner A., Saur J., 2017, ApJ, 835, 133


\bibitem[Servidio, Matthaeus \& Carbone (2008)]{ServidioETC08}
Servidio S., Matthaeus W. H., Carbone, V., 2008, 
Phys. Plasmas, 15, 042314

\bibitem[Servidio, Matthaeus \& Dmitruk (2008)]{ServidioMattheausDmitrukPRL08}
Servidio S., Matthaeus W. H., Dmitruk P., 2008, Phys. Rev. Lett., 100, 095005


\bibitem[\protect\citeauthoryear{Servidio et al.}{2015}]{ServidioETAL2015}
Servidio S. et al. 2015, J. Plasmas Phys., 81, 325810107


\bibitem[\protect\citeauthoryear{Smith et al.}{2006}]{Smith06}
Smith C.W. et al., 2006, ApJ,
645, L85

\bibitem[Stribling \& Matthaeus (1991)]{StriblingMatthaeusPoF91}
Stribling T., Matthaeus W. H., 1991, Phys. Fluids B, 3, 1848

\bibitem[{Telloni \& Bruno (2016)}]{TelloniBrunoMNRAS16}
Telloni D., Bruno R., 2016, MNRAS, 463, L79

\bibitem[Waleffe (1992)]{W92}
Waleffe F. 1992, 
Phys. Fluids A, 4, 350

\bibitem[Wicks et al. (2010)]{WicksHorburyChenSchekochihinMNRASL2010}
Wicks R.T. et al., 2010, MNRAS, 407, L31

\bibitem[Woltjer (1959)]{Woltjer59}
Woltjer L., 1959, Proc. Nat. Acad. Sci., 45, 769

\bibitem[\protect\citeauthoryear{Zhu}{2014}]{ZhuPoF14}
Zhu J.-Z., 2014, Phys. Fluids, 26, 055109

\bibitem[\protect\citeauthoryear{Zhu}{2016}]{ZhuJFM16}
Zhu J.-Z., 2016, J. Fluid Mech., 787, 440

\bibitem[\protect\citeauthoryear{Zhu}{2017}]{ZhuExact}
Zhu J.-Z. 2017, Revision of arXiv:1407.8404

\bibitem[Zhu \& Hammett(2010)]{gkaeOLD}
Zhu J.-Z., Hammett G. W., 2010, 
Phys. Plasmas, 17, 122307-1

\bibitem[Zhu, Yang \& Zhu(2014)]{hydrochirality}
Zhu J.-Z., Yang W., Zhu G.-Y., 2014, 
J. Fluid Mech., 739, 479



\end{thebibliography}

\label{lastpage}

\end{document}